# Isolating the Foreground of the X-Ray Background*

A. Refregier[1], D.J. Helfand[1], and R.G. McMahon[2]

[1] Columbia Astrophysics Laboratory, 538 W. 120th Street, New York, NY 10027, USA.
[2] Institute of Astronomy, Madingley Road, Cambridge CB3 0HA, UK.

**Abstract.** The foreground component of the X-ray Background (XRB) can be studied by cross-correlating its intensity with galaxy catalogs. We computed the two-point cross-correlation function $W_{xg}$ between 67 *Einstein* X-ray fields and the APM galaxy catalog. We detected a signal with an amplitude of $W_{xg} = .04 \pm .01$, significant at the $3.5\sigma$ level. The reality of this signal was tested with a series of control data sets. We discuss how this signal can be used to constrain models of the XRB.

## 1. Introduction

One approach to study the cosmic X-ray Background (XRB) is to probe its foreground component by cross-correlation with galaxy catalogs. This method provides information not only on the contribution of the catalogued galaxies themselves to the XRB but also on the relation of the XRB to local large scale structure and clustering.

A recent series of studies has focused on the two-point cross-correlation function $W_{xg}$ between the hard XRB ($\sim 2 - 10$ keV) and various optical and infra-red galaxy catalogs (Jahoda et al. 1991; Lahav et al. 1993; Miyaji et al. 1994; Carrera et al. 1995). A positive correlation signal was detected. When extrapolated to $z \sim 5$, this signal would indicate that 10-30% of the XRB is due to sources associated with galaxies. More recently, Roche et al. (1995) have found a positive cross-correlation signal between three soft ROSAT-PSPC fields (.5-2.0 keV) and galaxies down to $B \sim 23$. Their use of an imaging X-ray telescope allowed them to probe much smaller angular scales ($\sim 15''$) than the aforementioned studies in the hard band ($> 1\text{deg}^2$). However, their statistics were limited by the small number of fields considered. Treyer & Lahav (1995) have concluded from the soft results that the faint galaxies could contribute about 22% of the XRB in this band.

In order to test these results at intermediate angular scales ($\sim 1'$) and with improved statistics, we computed $W_{xg}$ for 67 *Einstein*-IPC fields with the APM northern galaxy catalog ($E < 19$). The *Einstein* energy-band (.1-2.7 keV) allows flexibility for testing the energy dependence of the correlation. In addition, a large enough signal would allow us to consider several magnitude slices for the galaxies, thereby testing, at least crudely, the redshift dependence of the correlation. In this paper, we focus on the zero-lag results and on the controls we have performed to test their significance. The current analysis along with non-zero lag measurements will be presented in greater detail elsewhere (Refregier et al. 1995).

## 2. Data

The X-ray data consisted of 67 *Einstein*-IPC fields in which sources were excised down to $3.5\sigma$. Source excision and flat fielding procedures are discussed in Wu et al. (1991) and Hamilton et al. (1991). We chose the fields to be at high Galactic latitude ($|b| > 30^0$ and to have exposure times greater than 6000 sec. We considered two energy bands: "soft" (.1-.56 keV) and "hard" (.56-2.70 keV). The "hard" band should be viewed as the real data whereas the "soft" band, which is dominated by emission from the Milky Way, should be considered as a control data set. Another control data set was constructed by including exposures with high solar contamination (VG=2,3); the "real" data set contained only low solar contamination (VG=1,2) exposures. Each one-square-degree field was subdivided in $64''$ pixels which contain most of the IPC point spread function ($\sim 1'$ at $2\sigma$)

The APM northern catalog was derived from scans of Palomar Observatory Sky Survey plates (Irwin & McMahon 1992; Irwin et al. 1994.) In order to maximize the performance of the automatic image classifier, we restricted our analysis to the red plate objects with E-magnitudes between 13.5 and 19. This sample contained about 530 galaxies per square degree. To avoid complications at plate boundaries, we chose the field centers to lie further than $45'$ from the plate edges. The stars found in the same region of the sky were used as an additional control data set. The galaxies (and the stars) were also binned into $64''$ pixels.

## 3. Procedure

Our main task was to compute the zero-lag two-point cross-correlation function $W_{xg}$ between the X-ray fields and the galaxies. This function gives a measure of the probability above random to find an excess in the XRB intensity at the location of a catalogued galaxy. A finite-cell estimator of $W_{xg}$ is given by

$$W_{xg} = \frac{n_{cell} \sum_i N_i I_i}{(\sum_i N_i)(\sum_i I_i)} - 1, \quad (1)$$

where $I_i$ and $N_i$ are the X-ray intensity and the number of galaxies in the $i^{th}$ cell, and $n_{cell}$ is the total number of cells (Jahoda et al. 1991). The cell size was varied by binning adjacent pixels.

Since both the X-ray and the galaxy data were subject to field-to-field variations (due differences in particle background contamination, plate sensitivity, etc), the 67 fields could not be treated as a single data set. Instead, we computed $W_{xg}$ for each field separately, and computed the mean and $1\sigma$ uncertainty estimate ($\delta W_{xg} \simeq \sigma_{W_{xg}}/\sqrt{67}$ where $\sigma_{W_{xg}}$ is the rms standard deviation.) The calculation was performed for the real data set and for the aforementioned control data sets. In addition, we tested the statistical significance of our results by producing 400 scrambled realizations of the real X-ray-galaxy field pairs.

## 4. Results

Figure 1 shows $W_{xg}$ with cell sizes ranging from 1 to 4 pixels for the different data sets. The real data set (hard-galaxies) produces a positive correlation of $W_{xg} = .04 \pm .01$, with a significance of $3.5\sigma$ (for a cell size of $64''$). Although the measurements at different cell sizes are not independent, their comparison gives a measure of the robustness of the signal. As expected, the soft and solar-contaminated control sets yield lower correlation signals. The hard-stars values for $W_{xg}$ are very close to zero. It is important to note that although the scrambled values are close to zero on average, their uncertainties ($\delta W_{xg} \simeq .01$) are not negligible. This gives a measure of the statistical uncertainty in the correlation signal.

## 5. Conclusions

We have detected a significant correlation signal between *Einstein* XRB intensity and APM galaxies. The signal passes our control tests. We emphasize the importance of good statistics and of the scrambled data set in obtaining an estimate for the statistical uncertainty involved.

A detailed interpretation of the signal is beyond the scope of this paper. We simply note that, in general, a correlation could result from three effects: (1) X-ray emission from the catalogued galaxies themselves, (2) a spatial correlation between the galaxies and the X-ray sources which form the XRB, or (3) X-ray emission from the hot gas in clusters and groups of galaxies. The first two effects have been referred to as "Poisson" and "clustering" terms in the literature (eg. Lahav et al. 1993). The last effect has been ignored before, but it can be significant. These three contributions have different intrinsic angular scales and can perhaps be decoupled by a non-zero-lag analysis. We also note that the point-spread function must be taken into account while dealing with imaging X-ray instruments. A discussion of these issues will be presented in Refregier et al. (1995).

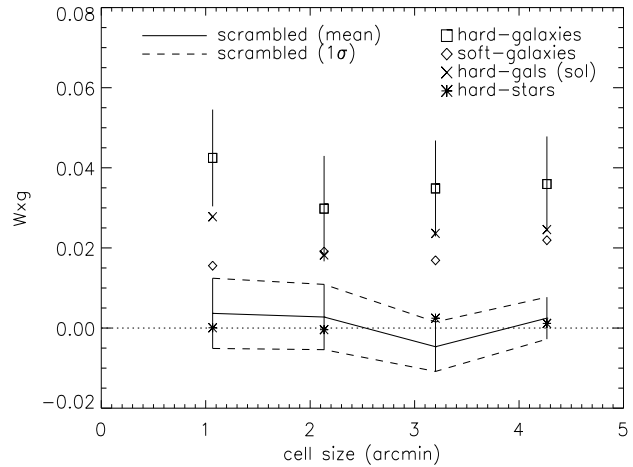

**Fig. 1.** $W_{xg}$ as a function of the cell size for the real (hard-galaxies) and control data sets (the soft, predominantly Galactic emission with the galaxies, the hard emission with high solar contamination with the galaxies, the hard-band emission with the stars and the hard-galaxies scrambled field pairs.) For clarity, the $1\sigma$ uncertainties are only shown for the real and scrambled sets.

*Acknowledgements.* We thank E. Moran for customizing the *Einstein* archive software and S. Maddox and M. Irwin for their help with the APM catalog. We also thank O. Lahav, M. Treyer, A. Blanchard, and R. Pilla for useful discussions. This work was supported by grant NAGW2507 from NASA. This paper is Contribution Number 585 of the Columbia Astrophysics Laboratory.